\documentclass[final,english]{bullsrsl}
\usepackage[latin1]{inputenc}
\usepackage[T1]{fontenc}
\usepackage{natbib} 
\usepackage{graphicx,amsmath}

\begin{document}
\title{On the Constancy of Binary Star Fractions in Local Star Clusters}

\author[affil={1,2}, corresponding]{Priya}{Hasan}
\author[affil={1}]{Mohammed}{Saifuddin}
\affiliation[1]{Department of Physics, Maulana Azad National Urdu University, Hyderabad, India 500032}
\affiliation[2]{Inter-University Centre for Astronomy and Astrophysics, Post Bag 4, Ganeshkhind, Savitribai Phule Pune University Campus, Pune 411007, Maharashtra, India.}
\correspondance{priya.hasan@gmail.com}
\date{16 February 2026}
\maketitle

\begin{abstract}
We investigate the unresolved binary-star fraction in a complete sample of 376 open clusters located within 1 kpc of the Sun using Gaia DR3 and ASteCA based synthetic colour magnitude diagram modelling. The binary fraction is a global photometric quantity, sensitive primarily to unresolved binaries that produce measurable broadening above the single-star main sequence. We find that the median binary fraction of the sample is nearly constant, $0.29\pm 0.03$, with no clear dependence on cluster age, core radius, limiting radius or median mass. This apparent constancy suggests that the global binary content of nearby open clusters is regulated by competing dynamical processes, including disruption of soft binaries, survival and hardening of close binaries, mass segregation, evaporation of single stars, and long-term cluster dissolution. The result is consistent with theoretical expectations and previous observational studies indicating that cluster binary populations can evolve internally while maintaining a relatively stable global binary fraction. Our findings provide an observational constraint for models of open-cluster formation and dynamical evolution and highlight the need for future work combining photometric, astrometric, spectroscopic, and N-body approaches to distinguish between primordial and dynamically formed binary populations.
\end{abstract}

\keywords{CMD, star clusters, Gaia DR3, stellar evolution}

\section{Introduction}

The binary fraction is defined as the ratio of the number of binary stellar systems to the total number of stellar systems in a star cluster. 
The binary stellar systems are probabilistically identified as unresolved stars and estimated on the basis of the analysis of the number of stars displaced in the main sequence through an  efficient statistical approach corrected for selection biases \citep{Sollima2007}. 

It is known that binary stars are common, nearly half of solar type stars are in pairs, with half of those being separated by more than 100 au, and hence vulnerable to disruption \citep{2013ARA&A..51..269D}. If these systems are easily disrupted by interactions, 
we should find more of them in younger populations, like in open clusters, with the fraction declining with age \citep{offner2022originevolutionmultiplestar}.

Data from the ESA Gaia mission \citep{2023A&A...674A...1G} is invaluable in the study of open clusters, as it provides homogeneous  6D information on stellar parallax, coordinates, proper motions and radial velocities. 
Stars in binaries, even wide binaries, should be moving together through space, and the precision of the Gaia measurements is such that associations stand out from the collective motion of the cluster. Hence, Gaia data can be used very effectively to identify members in contrast to earlier methods that were less reliable.   

The binary fraction is crucial to our understanding of stellar and dynamical evolution, as single stellar evolution is better understood than binary evolution. Observations of embedded protostars and young stellar populations show that multiplicity is a common outcome of star formation, although the multiplicity fraction depends strongly on stellar mass, separation range, and environment. The primordial binary or multiple-star fraction is  expected to be high, particularly for solar-type and higher-mass stars \citep{2013ARA&A..51..269D}. This high initial multiplicity is generally understood as a consequence of fragmentation during star formation, including turbulent fragmentation of molecular clouds, fragmentation of dense cores, and disk fragmentation.

To understand their evolution, we need to understand the properties of binaries (their separation distribution, mass ratios, and eccentricity), which provide critical tests for star formation theories. For example, does the fragmentation process prefer creating tight binaries or wide ones? And then how do they evolve after they are formed?

The following factors  
influence the binary fraction:
\begin{itemize}
\item 
Cluster density: Denser clusters have higher rates of stellar collisions, which can dissolve binary systems. This can lead to lower binary fractions. 
\item 
Cluster age: The binary fraction generally decreases over time as binaries are disrupted. 
\item 
Cluster mass: For some clusters, the binary fraction is not correlated with mass, especially for massive clusters. However, studies show that a higher cluster mass can lead to a higher fraction of ejected binaries, depending on the initial cluster density.

\item 
Metallicity: 
Metallicity affects the physical conditions under which stars form, and low metallicity may favour the formation of close binaries. Young clusters often have a narrow, metal-rich distribution.

\item 
Environment:  Binary fractions are often higher in lower-density environments like filamentary and fractal clusters compared to denser halo-type clusters. 
\item 
Formation process: Binary fraction can be affected by the cluster's formation process, particularly by the hierarchical assembly of subclusters and star formation rates. 

\end{itemize}

\section{Data and ASteCA}
\begin{figure}
\centering
\includegraphics[width=0.7\linewidth,angle =270]{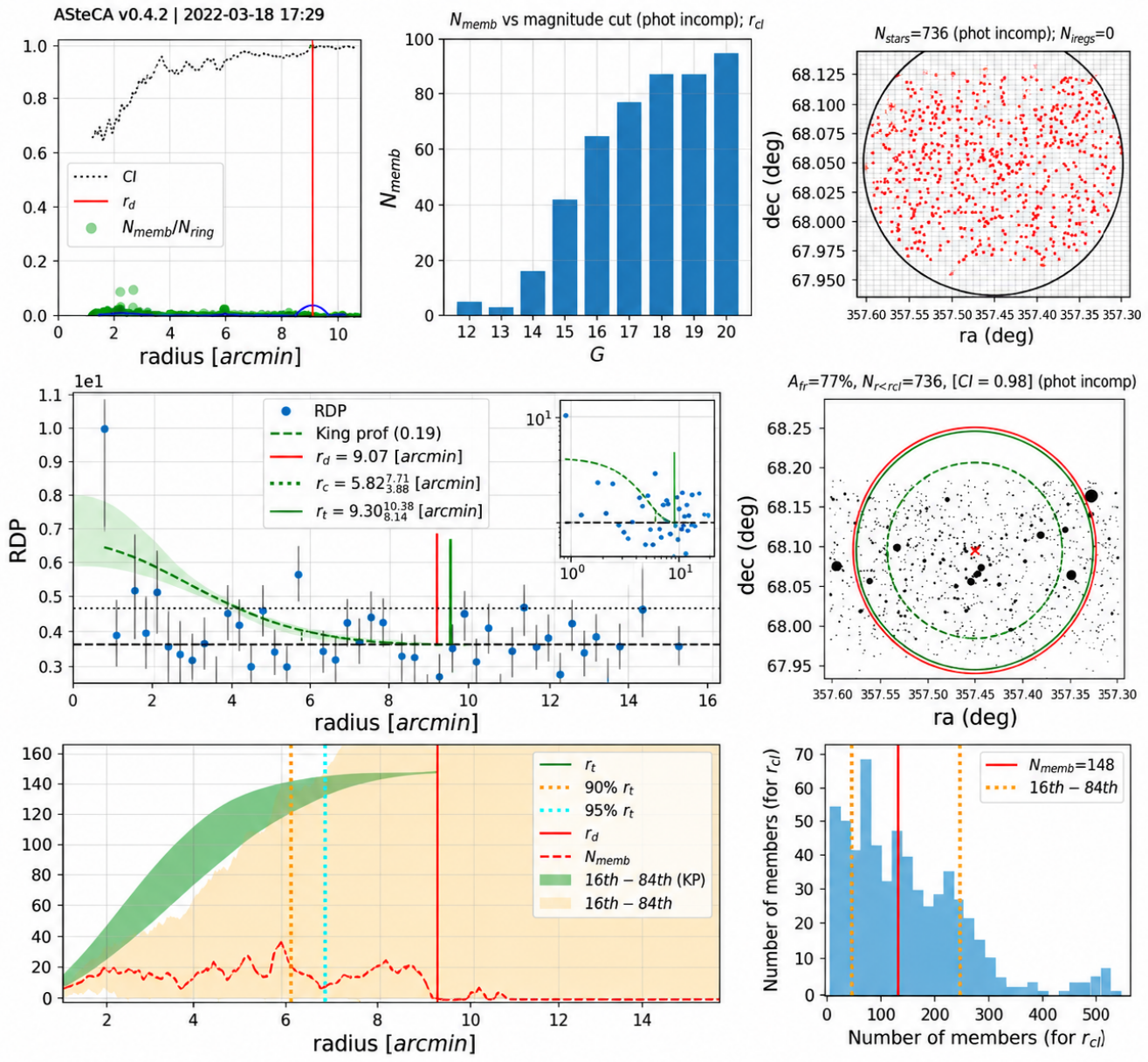}
\bigskip
\begin{minipage}{12cm}
\caption{ASteCA steps to characterize the cluster: The upper-left panel
determines the observational cluster radius, $r_{\mathrm{cl}}$, from the variation of the field-subtracted member fraction and contamination index, with radius marked by the red vertical line. The upper-middle panel shows how the estimated number
of members changes as progressively fainter stars are included, thereby revealing the effect
of photometric incompleteness. The upper-right panel displays the complete stellar field and the
adopted cluster boundary. The middle-left panel presents the radial-density profile  and the fitted King model, from which the core radius, $r_{\mathrm{c}}$, and tidal radius, $r_{\mathrm{t}}$
are obtained. The middle-right panel shows the enlarged spatial distribution, including the estimated centre, observational boundary, and fitted King core and tidal regions. The lower-left
panel compares the cumulative field-subtracted member count with that predicted by the King
model as a function of radius, while the lower-right panel gives the distribution of possible
member counts obtained by propagating the uncertainty in the field density. A more detailed explanation is provided in the text.}
\label{fig1}
\end{minipage}

\end{figure}

The Automated Stellar Cluster Analysis (ASteCA) \citep{ast15} is a specialized software package designed for the quantitative analysis of resolved stellar populations in dense star clusters. In this case ASteCA  makes use of Gaia DR3 to obtain precise and objective values for a given cluster center coordinates, radius, luminosity function and integrated color magnitude. It incorporates a Bayesian field star decontamination algorithm capable of assigning membership probabilities using photometric data alone. It also provides accurate estimates for a cluster's metallicity, age, extinction and distance values along with its uncertainties. Figure \ref{fig1} shows the ASteCA analysis for NGC~7762. 

The analysis of binary systems in ASteCA involves three distinct
stages: the generation of synthetic binaries, the fitting of the cluster
parameters, and the identification of probable binaries in the observed
sample. During the synthetic-generation stage, the probability that a stellar system is binary is, by default, a function of the primary-star mass
(\url{https://asteca.readthedocs.io/en/latest/contents/synthetic_mod.html#intrinsic-parameters}). The secondary-star mass is also drawn from a distribution that depends on the primary mass (\url{https://asteca.readthedocs.io/en/latest/contents/synthetic_mod.html#binary-systems}). Together, these prescriptions determine the binary population included in each synthetic cluster. Every binary system becomes a single (unresolved) datapoint in the synthetic CMD, irrespective of the binary orbital parameters, such as separation, eccentricity, inclination, or orbital phase. 

To estimate the fundamental parameters of an observed cluster,
ASteCA generates and evaluates thousands of synthetic clusters. The
observed and synthetic colour--magnitude diagrams (CMDs) are represented as
two-dimensional histograms and compared using a statistical likelihood. The
default objective function is the Poisson likelihood ratio
(\url{https://asteca.readthedocs.io/en/latest/contents/likelihood_mod.html#poisson-likelihood-ratio}),
although the user may implement an alternative function. This comparison is
used to constrain all fitted fundamental parameters, including age,
metallicity, extinction, distance, total mass, and binary fraction, rather
than binarity alone.

After the best-fitting parameters have been obtained, ASteCA performs
a post-processing analysis to identify probable binary systems in the
observed data
(\url{https://asteca.readthedocs.io/en/latest/contents/synthetic_mod.html#post-process-parameters}).

An observed source is classified probabilistically according to
whether its position in photometric space is closer to the distribution of
synthetic binary systems or to that of synthetic single stars. By repeatedly
examining the neighbouring synthetic systems, ASteCA assigns each
observed source a probability of belonging to the synthetic binary population.  `Binary' in this context refers to a source whose photometry is consistent with an unresolved binary sequence in the CMD, not to a binary that has been spatially or spectroscopically resolved.

\begin{figure}[!htbp]
\centering
\includegraphics[width=0.6\linewidth, angle =270]{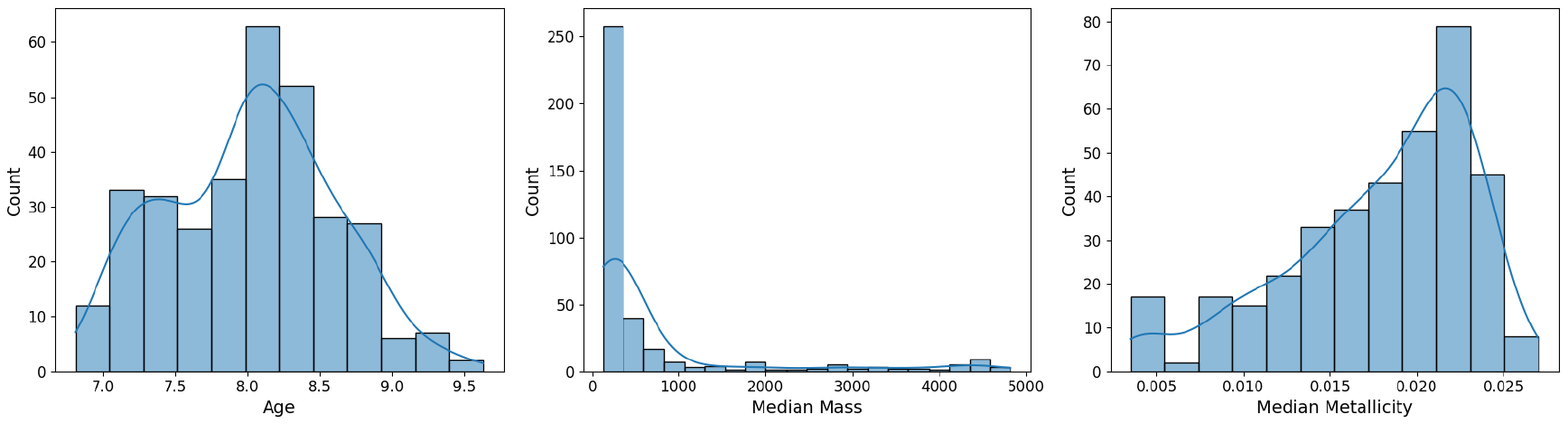}
\bigskip
\begin{minipage}{12cm}
\caption{Age, Mass and Metallicity distribution of our sample clusters}
\end{minipage}
\label{s2}
\end{figure}

For this work, we selected a sample of 376 open clusters -- {all clusters within 1000 pc} of the Sun from the catalog of \citet{2020A&A...640A...1C} and ran ASteCA for the sample with Age, Mass and Metallicity distribution described in Fig.~\ref{s2}.

\section{Results}
\begin{figure}[!htbp]
\centering
\includegraphics[width=0.45\linewidth]{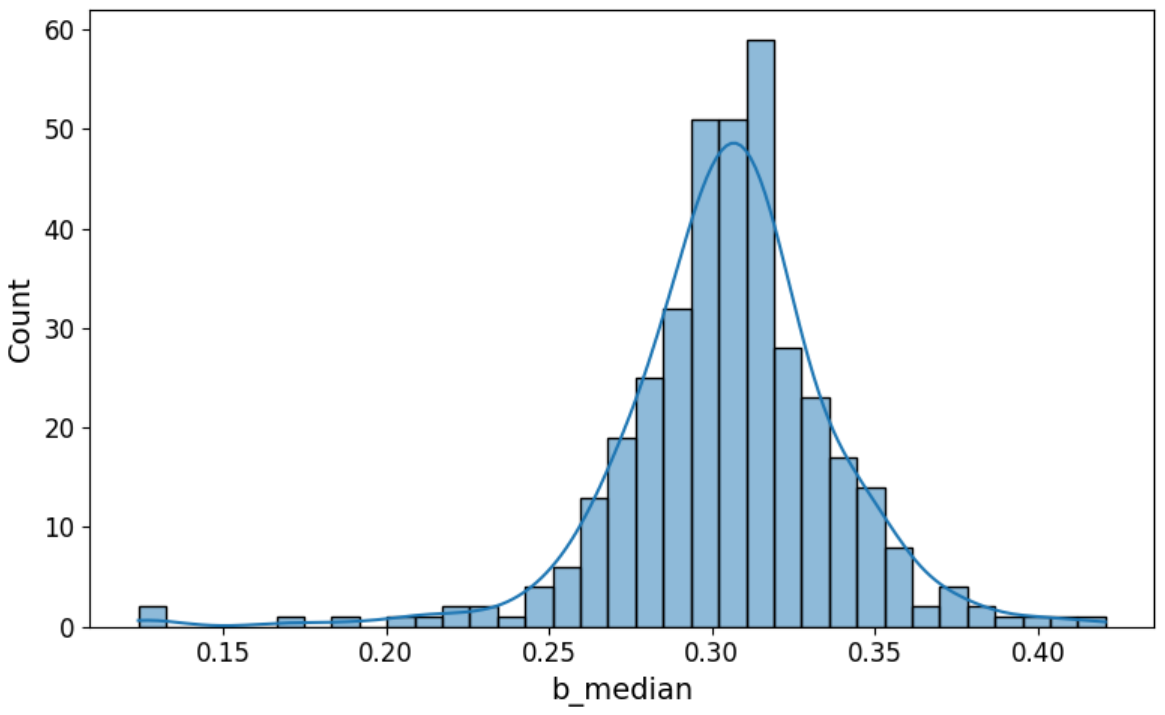} \includegraphics[width=0.45\linewidth]{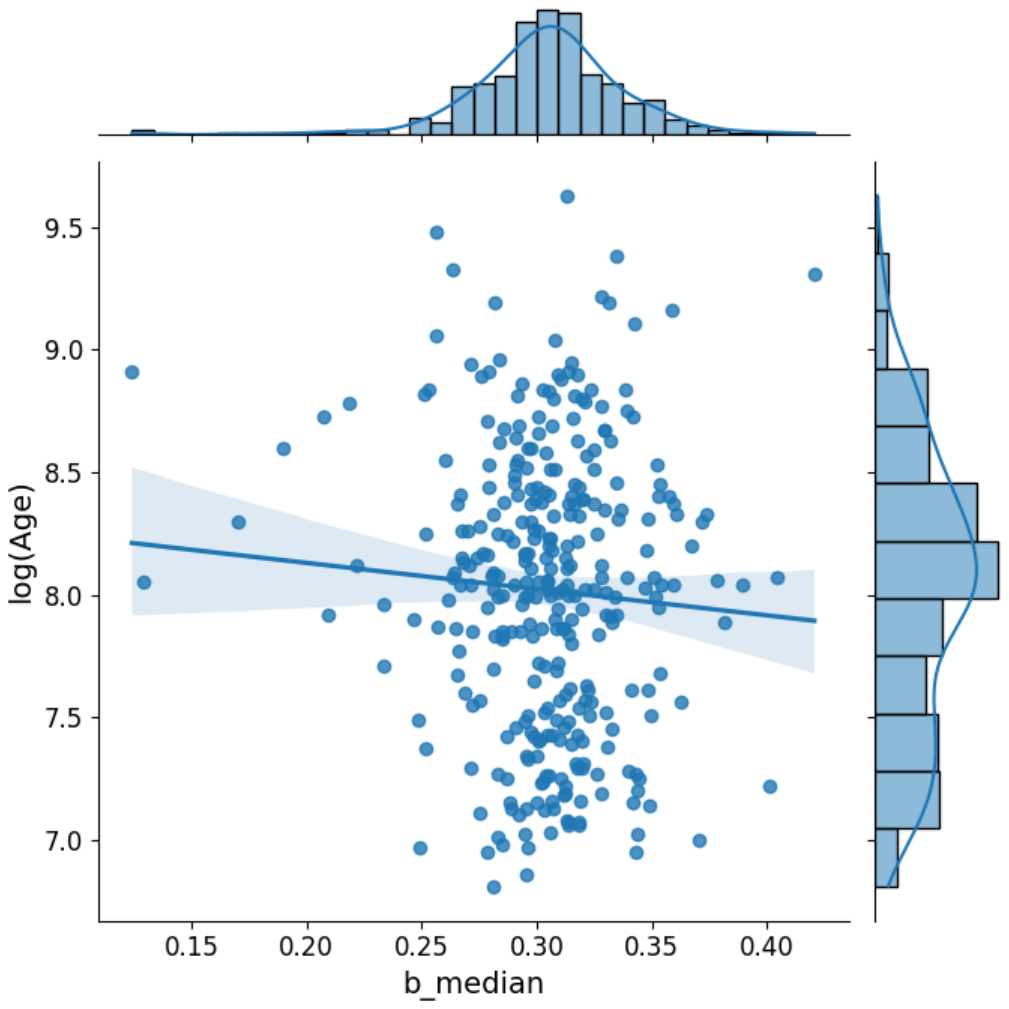}
\bigskip
\begin{minipage}{12cm}
\caption{Left: Binary Fraction of our sample clusters Right: Binary fraction vs log(age) AgeNN of clusters}
\label{bf}
\end{minipage}

\end{figure}

We find that the binary fraction seems to be almost constant $0.29 \pm0.03$ for all clusters within 1000 pc of the Sun, irrespective of size, age and mass as shown in Fig. \ref{bf}.

\begin{figure}
\centering
\includegraphics[width=0.8\linewidth,angle =270]{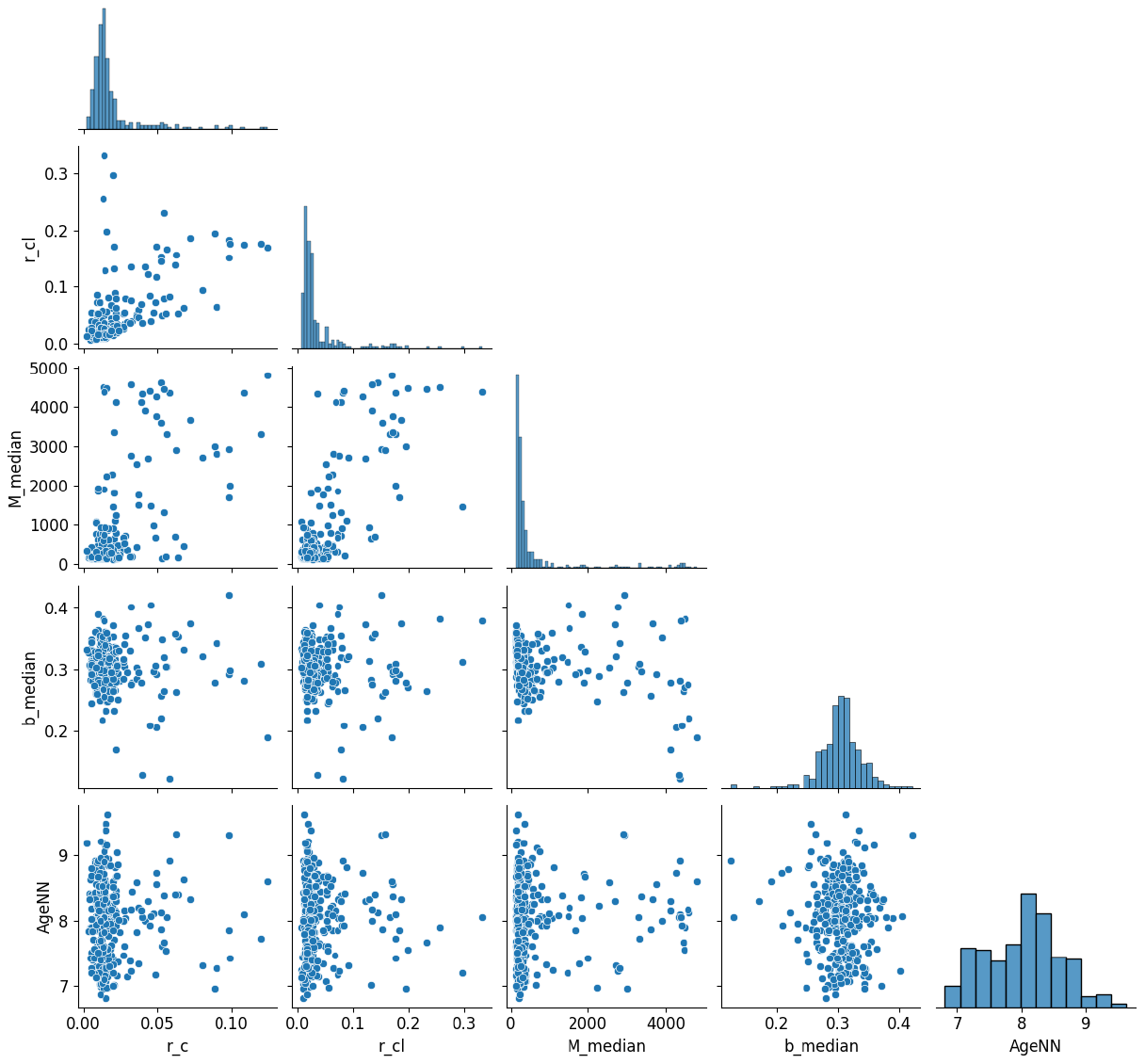}
\bigskip
\begin{minipage}{12cm}
\caption{Corner plots of parameters using results of ASteCA}
\label{fig3}
\end{minipage}

\end{figure}

Figure \ref{fig3} shows the corner plots for the parameters obtained from ASteCA. In the plot, we explore the inter-dependence of  the core radius ($r_{\mathrm{c}}$), cluster radius ($r_{\mathrm{cl}}$), the median mass ($M_{\mathrm{median}}$),  the median binary fraction ($b_{\mathrm{median}}$)  and log(Age) (AgeNN). However, as visible in the plot, there does not seem to be clear dependence of any of these factors on $b_{\mathrm{median}}$.  

\section{Discussion and Conclusions}

In this work, we analysed the unresolved binary-star population of 376 nearby open clusters located within 1 kpc of the Sun using Gaia DR3 data and ASteCA. The main result is that the binary fraction is approximately constant across the sample, with a median value of $0.29\pm0.03$. The corner plots shown in Fig. \ref{fig3} indicate no  dependence of the binary fraction on cluster age, core radius, limiting radius and median mass.

This is the photometric unresolved binary fraction, and not a complete census of all binary systems. In real cluster CMDs, the main sequence is broadened by several effects, including photometric uncertainties, field-star contamination, differential reddening, stellar rotation, and unresolved binaries. Unresolved binaries are particularly important because the combined light of two stars shifts the system above the single-star main sequence, producing an asymmetric broadening. The use of synthetic CMD fitting with ASteCA is therefore appropriate, since ASteCA estimates cluster parameters by comparing observed CMDs with synthetic stellar populations generated from theoretical isochrones \citep{ast15}.

The near constancy of the binary fraction  is also significant because binary populations are expected to evolve dynamically. Stellar multiplicity is a common outcome of star formation, but the multiplicity fraction depends strongly on stellar mass, separation range, and environment \citep{2013ARA&A..51..269D}. Thus, the present-day binary fraction of a cluster is not simply a primordial quantity; it reflects both the initial conditions of star formation and the later dynamical evolution of the cluster.

Cluster dynamics provides a natural explanation for why the global binary fraction may remain nearly stable. According to Heggie, hard binaries tend to become harder during stellar encounters, while soft binaries become softer and are more easily disrupted \cite{1975MNRAS.173..729H}. As a result, wide and weakly bound binaries are preferentially destroyed, whereas close binaries are more likely to survive. Dynamical evolution changes the separation and binding-energy distribution of binaries, even if the total global binary fraction changes only weakly. This interpretation is supported by N-body simulations. \citet{Hurley_2007} showed that the global binary fraction of a star cluster can remain close to its primordial value for much of the cluster lifetime, except near the final stages of dissolution. The core binary fraction can increase substantially with time because binaries are more massive than single stars on average and tend to sink toward the cluster centre through mass segregation. As a result, wide and weakly bound binaries are preferentially disrupted, whereas close, strongly bound binaries are more likely to survive. Even if wide binaries constitute only a small fraction of the total binary population, their selective destruction can substantially truncate the high-separation, low-binding-energy tail of the distribution without strongly changing the global binary fraction. Moreover, preferential evaporation of lower-mass single stars can partly compensate for the loss of binaries and may even increase the binary fraction among the remaining cluster members.

Photometric estimates of the binary fraction are most sensitive to unresolved binaries with mass ratios close to unity, where ($q=m_2/m_1\leq$1), because both components contribute appreciably to the combined luminosity and displace the system measurably from the single-star sequence in the colour--magnitude diagram. \citet{Fregeau_2009} also found that binary destruction, mass segregation, and preferential loss of single stars can partly balance each other, producing relatively stable global binary fractions. Therefore, the constancy found in the present work does not imply that binaries are dynamically unchanged; rather, it suggests that competing dynamical processes may regulate the global binary content.

Observational studies are consistent with this picture. \citet{elson98} found that the binary fraction in the young LMC cluster NGC 1818 increases toward the cluster centre, indicating that local binary fractions can vary strongly with radius. \citet{2012A&A...540A..16M} used HST observations of Galactic globular clusters, also found that binaries are generally more centrally concentrated than single main-sequence stars and that binary fraction is more strongly related to cluster mass or luminosity than to many other global parameters. These studies show that binary populations evolve internally, even when simple global trends with age, radius, or mass are weak.
The median value obtained in this study, $0.29 \pm0.03$, is also broadly consistent with recent measurements for Galactic open clusters. \citet{yalya24} estimated binary fractions of about 0.3--0.5 for FSR 866, NGC 1960, and Stock 2 using Gaia DR3 astrometry together with photometric and spectroscopic methods. This agreement suggests that binary fractions of a few tens of percent are typical for open clusters when measured using photometric or combined photometric-spectroscopic techniques.

The absence of a strong age dependence is particularly important. A simple expectation might be that older clusters should show lower binary fractions because binaries are disrupted over time. However, this expectation is incomplete. Soft binaries are disrupted, hard binaries survive and harden, binaries migrate toward the cluster centre through mass segregation, and single stars may be preferentially lost through evaporation. These effects can compensate for one another, producing an approximately constant global binary fraction even though individual binary systems continue to evolve.

Several limitations in this work should be considered. Photometric binary fractions are most sensitive to unresolved binaries with relatively high mass ratios; low-mass-ratio binaries lie close to the single-star main sequence and are difficult to identify. In addition, CMD broadening can arise from effects other than binarity, including photometric errors, extinction variations, rotation, and residual field contamination. The sample is also limited to clusters within 1 kpc, where Gaia data quality is high but selection effects may still favour clusters with clearer CMDs and more reliable memberships.

Overall, this study suggests that nearby open clusters share a remarkably uniform global unresolved binary fraction of about 30\%. This constancy provides an important observational constraint for models of cluster formation and dissolution. Any successful model of open-cluster evolution should reproduce not only the presence of binary systems, but also the apparent stability of the global binary fraction across the sampled range of ages, sizes, masses, and Galactic environments. Future work should test this result by grouping clusters according to dynamical age, relaxation time, density, and Galactocentric distance, and by combining photometric binary fractions with Gaia astrometry, radial-velocity monitoring, high-resolution imaging, and direct N-body simulations.

\begin{acknowledgments}
The authors would like to thank the referee for valuable comments, which helped improve the paper.
This work has made use of data from the European Space Agency (ESA) mission
{\it Gaia} (\url{https://www.cosmos.esa.int/gaia}), processed by the {\it Gaia}
Data Processing and Analysis Consortium (DPAC,
\url{https://www.cosmos.esa.int/web/gaia/dpac/consortium}). Funding for the DPAC
has been provided by national institutions, in particular the institutions
participating in the {\it Gaia} Multilateral Agreement.
\end{acknowledgments}

\begin{furtherinformation}
\begin{orcids}
\orcid{0000-0002-8156-6940}{Priya}{Hasan}
\end{orcids}

\begin{authorcontributions}
MS is a PhD student working on his thesis `Star Formation in the Gaia Era' and has been doing the analysis, plotting, coding, and writing. PH is the thesis supervisor involved in conceptualization, writing, and guiding the student. 
\end{authorcontributions}

\begin{conflictsofinterest}
The authors declare that there is no conflict of interest.
\end{conflictsofinterest}
\end{furtherinformation}

\bibliographystyle{bullsrsl-en}

\bibliography{ref}

\end{document}